\documentclass[prl,amsmath,twocolumn, showpacs, superscriptaddress,10pt]{revtex4-1}

\usepackage{amsmath}
\usepackage{hyperref}
\usepackage{graphicx}

\begin{document}

\title{First order transition for the optimal search time of L\'evy flights with resetting}

\author{Lukasz \surname{Kusmierz}}
\affiliation{Institute of Physics, UJ, Reymonta 4, 30–059 Krakow, Poland}
\affiliation{Institute of Automatics, AGH, Al. Mickiewicza 30, 30-059, Krakow, Poland}
\author{Satya N. \surname{Majumdar}}
\affiliation{Univ. Paris-Sud, CNRS, LPTMS, UMR 8626, Orsay F-01405, France}
\author{Sanjib \surname{Sabhapandit}}
\affiliation{Raman Research Institute, Bangalore 560080, India}
\author{Gr\'egory \surname{Schehr}}
\affiliation{Univ. Paris-Sud, CNRS, LPTMS, UMR 8626, Orsay F-01405, France}

\begin{abstract}
We study analytically an intermittent search process in one dimension. There is an immobile target at the origin and a searcher 
undergoes a discrete time jump process starting at $x_0\geq0$, where successive jumps are drawn independently from an arbitrary jump distribution $f(\eta)$. In addition, with a probability $0\leq r \leq1$ the position of the searcher is reset to its initial position $x_0$. The efficiency of the search strategy is characterized by the mean time to find the target, i.e., the mean first passage time (MFPT) to the origin. For arbitrary jump distribution $f(\eta)$, initial position $x_0$ and resetting probability $r$, we compute analytically the MFPT. For the heavy-tailed L\'evy stable jump distribution characterized by the L\'evy index $0<\mu < 2$, we show that, for any given $x_0$, the MFPT has a global minimum in the $(\mu,r)$ plane at $(\mu^*(x_0),r^*(x_0))$. We find a remarkable first-order phase transition as $x_0$ crosses a critical value $x_0^*$ at which the optimal parameters change discontinuously. Our analytical results are in good agreement with numerical simulations.  
\end{abstract}

\maketitle


The study of search strategies has generated a tremendous interest over the last few years, as they have found a wide variety of applications in various areas of science. For instance, they play an important role in diffusion-controlled reactions \cite{BLMV11} -- with implications in the context of genomic transcription in cells \cite{BWH81} -- or in computer science~\cite{Lov96}, like in the quest of solution of hard optimization problem. More recently, search processes have been intensively studied in behavioral ecology \cite{Bel91}. In that context, searching for a target  is a crucial 
task for living beings to obtain food or find a shelter \cite{Bel91}. In this case, the survival of a species is conditioned, to a large extent, to 
the optimization of the search time. Hence the characterization of the efficiency of search algorithms has generated a huge interest
during the last few years, both experimentally \cite{Bel91, BBE90,HWQSS12} and theoretically \cite{VA96,VBHLRS99,BCMSV05,VLRS11,EM11a,EM11b}.

When studying animal movements during their search or foraging period, it has proven to be useful to model their outwardly unpredictable dynamics by random walks (RWs) \cite{BLMV11,VA96,VBHLRS99,BCMSV05,VLRS11,EM11a,EM11b}. The increasing number of experimental data for various animals~\cite{Bel91, BBE90,HWQSS12}, have stimulated the study of several search strategies based on RWs. 
In particular multiple scales RWs, where phases of local diffusion alternate with long range nonlocal moves, have been put forward as a viable and efficient search strategy. For instance, these nonlocal moves can be modeled by L\'evy flights \cite{VA96,VBHLRS99}, or by the so called ``intermittent'' RWs \cite{BCMSV05}.

Recently an intermittent strategy, where a locally diffusive searcher is reset randomly with a constant rate to its initial position, has been introduced 
and demonstrated to be rather efficient in searching a fixed target located at the origin in all dimensions~\cite{EM11a,EM11b,EMM13,WEM13,EM14,BS14}. In particular it was shown that the mean capture time of the target, a natural measure of the efficiency of the search process, is finite and becomes minimal
for an optimal choice of the resetting rate. Apart from the issue of search, this resetting move also drives the system to a
non-equilibrium stationary state which has been characterized fully both for a single
Brownian motion~\cite{EM11a,EM11b,EM14} and spatially extended systems including
fluctuating interfaces \cite{GMS14} or reaction-diffusion systems
\cite{DHP14} (in the latter case with a different resetting
procedure). In the last years, stochastic processes with random
restarts have also been used in computer science as a useful strategy
to optimize search algorithms in hard combinatorial problems as well as in simulated annealing
\cite{Lov96,MZ02}.

\begin{figure}
\includegraphics[width = 0.9\linewidth]{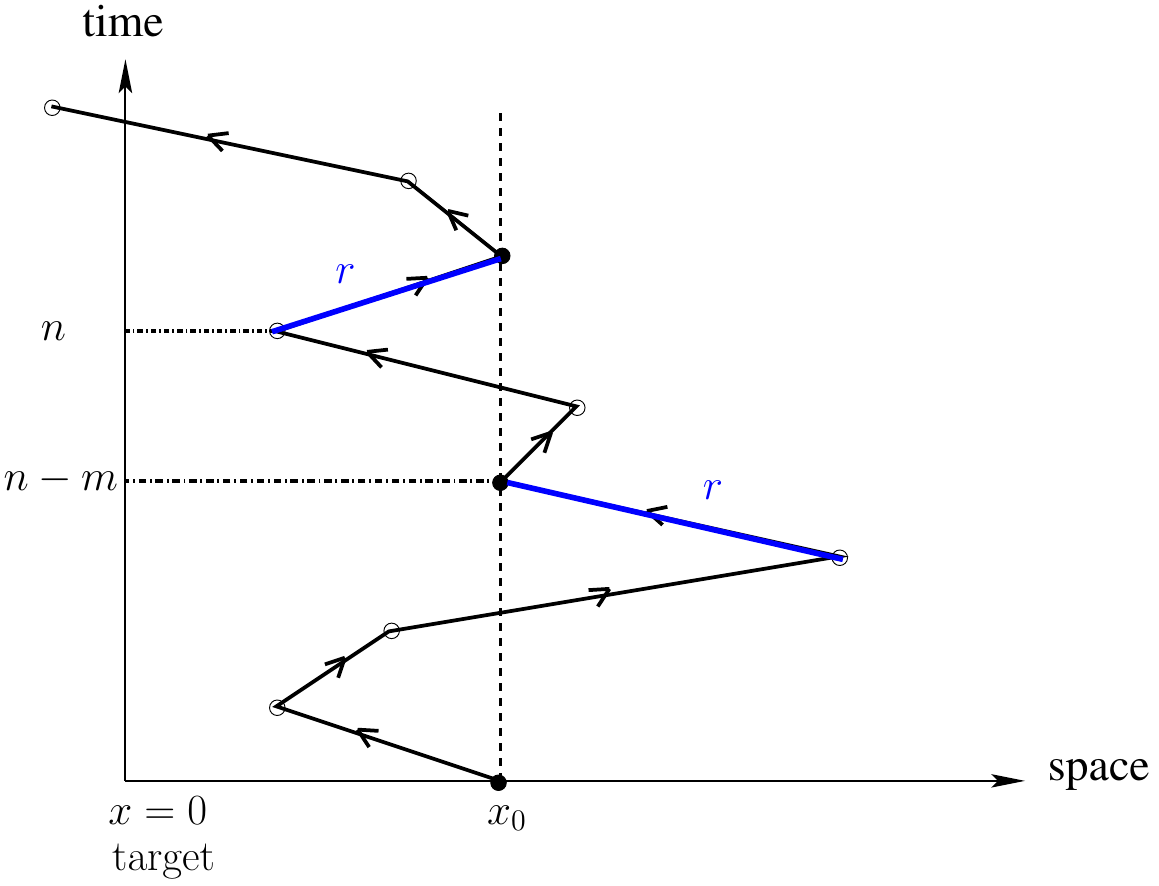}
\caption{Illustration of the search strategy which combines long jumps (L\'evy flights) and random resettings, with probability $r$, at the initial position $x_0$. Here, the search time, i.e. the first passage time in $0$ where the target is located is $T_{x_0}(\mu,r) = 9$ while there have been two resettings, at step $4$ and $7$. The integers $n$ and $m$ with $n=6$ and $m=2$ here illustrate the notations in the renewal equation in Eq. (\ref{Eq:renewal}). }\label{fig_intro}
\end{figure}

In all these situations discussed above, the local exploration process is typically diffusive. However animal movements on a local scale are
not always diffusive \cite{VA96,VBHLRS99} and the jump distribution between two successive positions may itself have heavy tails, such as in L\'evy 
flights. It is then natural to ask, for such jump processes with heavy tails, whether resetting to the initial position also makes the search of a target more efficient. In this Letter, we introduce a simple model that combines jump processes with heavy tails and random resetting to the initial position. Indeed
we demonstrate that resetting is an efficient search strategy even when the local moves are not Brownian, but are instead heavy tailed. 
In particular our analytical results demonstrate that this model has a rather rich behavior even in the simple one-dimensional setting, 
where it exhibits a rather surprising first-order phase transition.  

For simplicity, we define the model in one-dimension. Higher-dimensional generalizations of the model are straightforward. 
In our model, the searcher moves in discrete time on a line, starting from the initial position $x_0 \geq 0$. The target is located at the origin. 
At time step $n$, the current location $x_n$  
of the searcher is updated via the following rules~(see Fig. \ref{fig_intro}):
\begin{eqnarray}
x_{n}=
\begin{cases}
x_0\mbox{ with probability }r \\
x_{n-1} + \eta_n \mbox{ with probability }1-r,
\end{cases}
\label{Eq:model}
\end{eqnarray}
where $0< r < 1$ denotes the probability of a resetting event and the jump lengths $\eta_n$'s are independent and identically distributed
(i.i.d.) random variables each drawn from a probability distribution
function $f(\eta)$ with a heavy tail $f(\eta) \sim |\eta|^{-1-\mu}$ for large $|\eta|$, with a L\'evy index $0< \mu <2$. Here
we consider the class of L\'evy stable processes for which the Fourier transform of the jump distribution is given by $\hat f(k) =
\int_{-\infty}^{+\infty} e^{ik \eta} f(\eta) d \eta = e^{-|a k|^\mu}$ where $a$ sets the scale of the jumps (we set $a=1$ in
the following). The heavy
tail is reflected in the small $k$ behavior of $\hat f(k) \sim 1-|k|^\mu +\dotsb$ as $k\to 0$. The case $\mu = 2$ corresponds to ordinary random walks, while
$\mu < 2$ describes L\'evy flights where the jumps are typically very large \cite{feller}.
\begin{figure*}[ht]
\includegraphics[width = 0.3\linewidth]{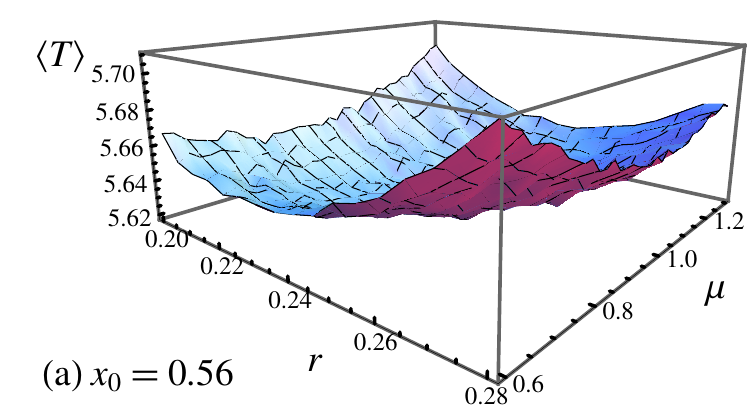}\includegraphics[width = 0.3\linewidth]{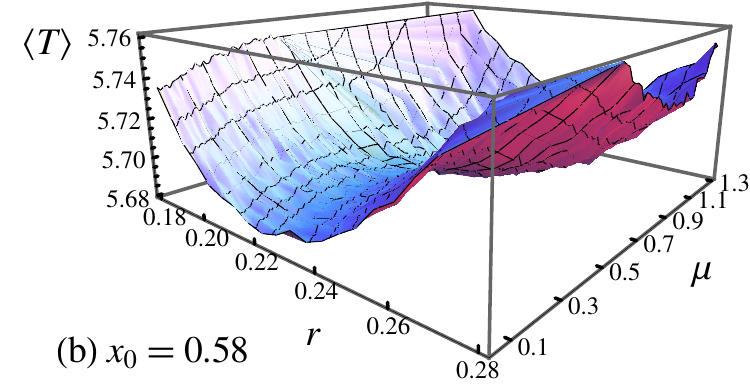}\includegraphics[width = 0.3\linewidth]{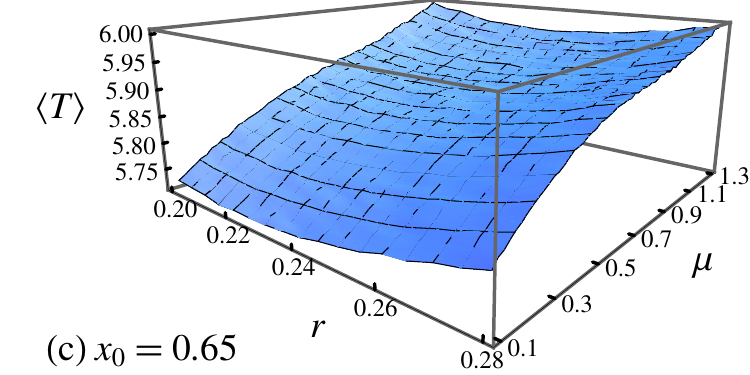}
\caption{2d plots of the average search time $\langle T_{x_0} (\mu,r)\rangle$, computed using numerical simulations, in the $(\mu,r)$ plane for different values of the initial position $x_0$: {\bf a)} $x_0=0.56< x_0^*$, {\bf b)} $x_0=x_0^*\simeq 0.58$ and {\bf c)} $x_0 = 0.65 > x_0^*$.}\label{fig_Tr}
\end{figure*}

\begin{figure}[h]
 \includegraphics[width = 1.\linewidth]{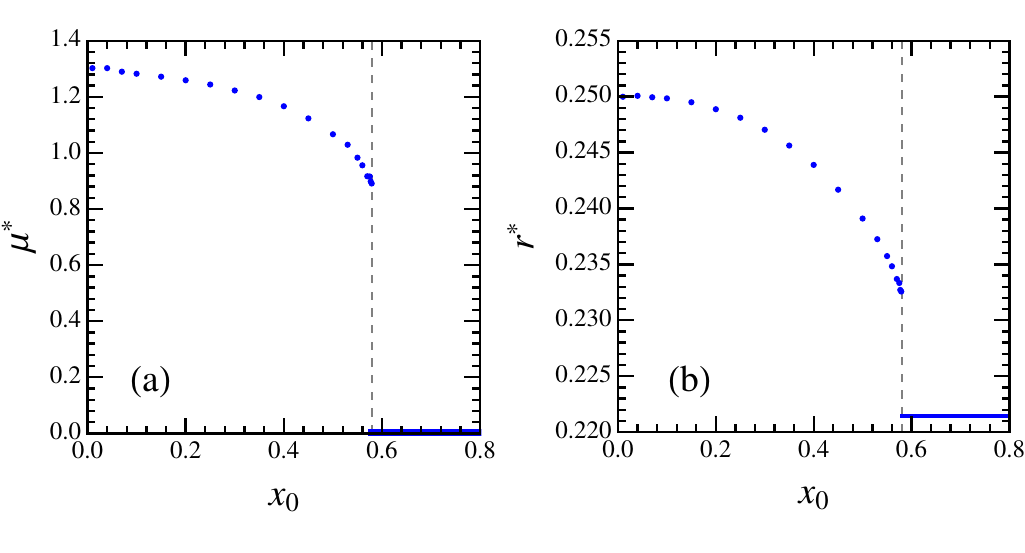}
 \caption{Plot of the optimal parameters $\mu^*(x_0)$ and $r^*(x_0)$, obtained from numerical simulations, as a function of $x_0$. Both of them exhibit a clear discontinuity for $x_0=x_0^* \simeq 0.58$, reminiscent of a first order phase transition.}\label{fig_mu_r}
 \end{figure}

In the following we consider the case of ``myopic search'' where the
search ends when the walker crosses the origin (the location of the immobile target) for the first time, (see Fig. \ref{fig_intro}). The efficiency of the search process is conveniently characterized by the average search time
$\langle T_{x_0}(\mu,r) \rangle$, which depends on $x_0$, $\mu$ and
$r$. In this Letter, we obtain an exact expression for $\langle
T_{x_0} (\mu,r)\rangle$ given in (\ref{Eq:Tr_main}).
For a fixed $x_0$, we then optimize $\langle T_{x_0}(\mu,r) \rangle$ with respect to the two parameters $\mu$ and $r$
and find the optimal parameters $\mu^*(x_0)$ and $r^*(x_0)$ as a function of $x_0$. Naively, one might have expected that the optimal
parameters are $\mu^* = r^* = 0$, independently of $x_0$. Instead we find, quite remarkably, that these optimal values $\mu^*(x_0)$ and $r^*(x_0)$
exhibit a rather rich and surprising behavior, as functions of
$x_0$. We show indeed that there exists a critical value $x_0^* \simeq 0.58$ (its value 
determined numerically) such that the optimal strategy depends
crucially on whether $x_0 > x_0^*$ or $x_0 < x_0^*$. When $x_0>x_0^*$,
the optimal parameters are independent of $x_0$, and are given by
\begin{subequations}
\begin{align}
&\mu^*(x_0>x_0^*) = 0 \;,\qquad r^*(x_0>x_0^*) = r_>^* \;,\\
\label{rstar_mu0}
& {\rm where}\; r_>^* = \frac{\sqrt{e-1}}{2} (\sqrt{e}-\sqrt{e-1})= 0.22145\ldots \;. 
\end{align}
\end{subequations}
On the other hand, for $x_0< x_0^*$, the optimal values $\mu^*(x_0)$
and $r^*(x_0)$ depend continuously on $x_0$, both of them being
monotonously decreasing functions of $x_0$. In particular, in the
limit where $x_0 \to 0^+$, we find
\begin{subequations}
\label{res_smallx0_intro}
\begin{eqnarray}
&&r^*(x_0 \to 0^+) = r^*_0 = 1/4 \;,  \\
&&\mu^*(x_0 \to 0^+) = \mu^*_0 = 1.2893\ldots \;,
\end{eqnarray} 
\end{subequations}
where $\mu^*_0$ is the solution of a transcendental equation given in~(\ref{Eq:muforsmallx0}). Moreover, we find that the optimal parameters $\mu^*(x_0)$ and $r^*(x_0)$ exhibit a discontinuity as $x_0$ crosses the value $x_0^*$ (see Fig. \ref{fig_mu_r}). This behavior is typically a characteristic of a first order transition at $x_0^*$.




In order to compute the mean search time, or the mean first passage time
(MFPT), $\langle T_{x_0}(\mu,r)\rangle$, to the origin ($x=0$), we
introduce the cumulative distribution function (CDF) $Q_{x_0}(r,n) =
{\rm Proba.}\,[T_{x_0}(\mu,r) \geq n]$. The CDF $Q_{x_0}(r,n)$ is thus
the survival probability, i.e., probability that the walker starting
from $x_0$ does not cross the origin up to step $n$ in presence of
resetting.  Obviously one has $\langle T_{x_0}(\mu,r) \rangle =
\sum_{n\geq0} Q_{x_0}(r,n)$. To compute $Q_{x_0}(r,n)$ we write a
recursion relation for this quantity, by using the fact that the
resetting dynamics in (\ref{Eq:model}) is Markovian. At a fixed time
step $n$, we denote by $m$ the number of steps elapsed since the last
resetting (see Fig. \ref{fig_intro}). Noting that the probability for
this latter event is $r (1-r)^{m}$ we have (see \cite{GMS14} for the
derivation of a similar equation in a continuous time setting)
\begin{eqnarray}
\nonumber Q_{x_0}(r,n)&=&\sum\limits_{m=0}^{n-1}r (1-r)^{m} Q_{x_0}(r,n-m-1)Q_{x_0}(0,m) \\
&+& (1-r)^n Q_{x_0}(0,n) \;,
\label{Eq:renewal}
\end{eqnarray}
where $Q_{x_0}(0,n)$ is the survival probability in the absence of resetting (i.e., $r=0$). The first term on the right hand side of Eq. (\ref{Eq:renewal}) accounts for the event where the last resetting before step $n$ takes place at step $n-m$ (see Fig. \ref{fig_intro}) with $0\leq m \leq n-1$. The evolution from step $n-m$ to step $n$ occurs without resetting and the survival probability during this period is $Q_{x_0}(0,m)$, 
while $Q_{x_0}(r,n-m-1)$ accounts for the survival probability from step $1$ to step $n-m-1$ in presence of resetting. 
The last term in Eq. (\ref{Eq:renewal}) corresponds to the case where there is no resetting event at all up to step $n$, which occurs with probability $(1-r)^n$. 

To solve Eq. (\ref{Eq:renewal}), we introduce its generating function $\tilde{Q}_{x_0}(r,z)=\sum_{n\geq0} Q_{x_0}(r,n)
z^n$. Multiplying both sides of Eq. (\ref{Eq:renewal}) by $z^n$ and
summing over $n$, we arrive at the result
\begin{equation}
\tilde{Q}_{x_0}(r,z)=\frac{\tilde{Q}_{x_0}(0,(1-r)z)}{1-r z \tilde{Q}_{x_0}(0,(1-r)z)} \;.
\label{Eq:Qr_result}
\end{equation} 
This formula (\ref{Eq:Qr_result}) relates the survival probability in presence of resetting ($r\geq 0$) to the one
without resetting ($r=0$). Fortunately, for any continuous and symmetric jump
distribution $f(\eta)$, the Laplace transform
(LT) of $\tilde Q_{x_0}(0,z)$ with respect to $x_0$ (the case of no resetting), can be explicitly
computed using the so-called Pollaczek-Spitzer
formula~\cite{Pol75,Spi56,Maj10}:
\begin{subequations}
\label{PS}
\begin{align}
&\int_0^\infty \tilde Q_{x_0}(0,z)e^{-\lambda x_0} \, {\mathrm d}x_0 = \frac{1}{\lambda\sqrt{1-z}} \varphi(z,\lambda) \;,  \\
&\varphi(z,\lambda) =  \exp{\left[-\frac{\lambda}{\pi} \int_0^{\infty} \frac{\mathrm d k}{\lambda^2+k^2}\ln{\left(1-z\hat{f}(k)\right)}\right]} \;.
\end{align}
\end{subequations}
Hence Eq. (\ref{Eq:Qr_result}) together with (\ref{PS}) allow us to
compute the CDF of the search time $T_{x_0}(\mu,r)$. Note that Eqs. (\ref{Eq:Qr_result}) and (\ref{PS})
are actually valid for arbitrary jump distributions $f(\eta)$, including in particular the L\'evy case in which we are interested. 

A useful characteristic of the full PDF of $T_{x_0}(\mu,r)$ is its first moment, on which we now focus. Noting the simple identity $\langle T_{x_0}(\mu,r)\rangle=\tilde{Q}_{x_0}(r,1)$, one obtains from~(\ref{Eq:Qr_result})
\begin{equation}
\langle T_{x_0}(\mu,r)\rangle=\tilde{Q}_{x_0}(r,1)=\frac{\tilde{Q}_{x_0}(0,1-r)}{1-r \tilde{Q}_{x_0}(0,1-r)} \;,
\label{Eq:Tr_main}
\end{equation}
where $\tilde{Q}_{x_0}(0,1-r)$ can, in principle, be computed from~(\ref{PS}). The first observation is that when $x_0 =  0$, the MFPT is totally independent of the jump distribution $f(\eta)$. Indeed, in this limit, $\tilde Q_{x_0=0}(0,z)$ is given by the Sparre Andersen theorem \cite{SA53,feller}, which states that $\tilde Q_{x_0=0}(0,z) = 1/\sqrt{1-z}$. Therefore, for $x_0 = 0$, one obtains a universal result
\begin{equation}
\langle T_{x_0=0}(\mu,r)\rangle=\frac{1}{\sqrt{r}-r} \;,
\label{Eq:Tzero}
\end{equation}
which is independent of $\mu$ and has a minimum at $r^*(x_0=0)  = 1/4$, where the minimal MFPT is~$T^*(x_0=0) = 4$. The question is: what happens when $x_0>0$? 

To get some insights for $x_0>0$, we first perform a small $x_0$ expansion of $\langle T_{x_0}(\mu,r)\rangle$. This requires the large $\lambda$
expansion of the LT of $\tilde Q_{x_0}(r,z)$ in Eq. (\ref{PS}). For the case of purely stable jumps, i.e. $\hat f(k) = e^{-|k|^\mu}$, this yields to lowest non-trivial order:
\begin{eqnarray}
\nonumber \langle T_{x_0}(\mu,r)\rangle=\frac{1}{\sqrt{r}(1-\sqrt{r})}-\frac{x_0}{\sqrt{r}(1-\sqrt{r})^2}\frac{1}{\pi}  \\
\times \int_0^{\infty}\ln{\left( 1 - (1-r) e^{-k^{\mu}}\right)}\mbox{d}k + O(x_0^2) \;.
\label{Eq:smallx0expansion}
\end{eqnarray}
We can now look for the optimal parameters $r^*(x_0)$ and $\mu^*(x_0)$
that minimize $\langle T_{x_0}(\mu,r)\rangle$ in
(\ref{Eq:smallx0expansion}), for a fixed (small) $x_0$. To lowest
order, we find $r^*(x_0) = 1/4 + O(x_0)$ while $\lim_{x_0\to
  0^+}\mu^*(x_0) = \mu^*_0$ where $\mu^*_0$ is the unique solution, on
the interval $(0,2)$, of the equation:
\begin{equation}
\int_0^{\infty}\frac{k^{\mu_0^*} \ln{k}}{\exp{(k^{\mu_0^*})}-{3}/{4}}\mbox{d}k=0 \;.
\label{Eq:muforsmallx0}
\end{equation}
Solving (\ref{Eq:muforsmallx0}) via Mathematica yields $\mu_0^* =
1.2893 \ldots$, as announced in (\ref{res_smallx0_intro}).
From (\ref{Eq:smallx0expansion}), one finds that the optimal MFPT is
given by $T^*(x_0) = \langle T_{x_0}(\mu^*,r^*)\rangle = 4 + {\cal O}(x_0)$.  This perturbative
calculation for small $x_0$ demonstrates the non-trivial fact that,
for small $x_0$, there exists a non-trivial optimal set of parameters
$(r^*(x_0), \mu^*(x_0))$ given in (\ref{res_smallx0_intro}). This leading order perturbation theory
can, in principle, be extended to higher orders in $x_0$. 

To proceed beyond the perturbative calculation presented above, we perform numerical simulations of the resetting dynamics in
(\ref{Eq:model}). For a given $x_0$, we compute numerically $\langle T_{x_0}(\mu,r) \rangle$ by sampling $10^7$ to $9\times 10^7$ (depending on $x_0$)
independent realizations of the
resetting dynamics (\ref{Eq:model}), for different values of the
parameters $r$ and $\mu$. In Fig. \ref{fig_Tr}, we show 
$\langle T_{x_0}(\mu,r) \rangle$ as a function of $\mu$ and $r$ for
three different values of $x_0$. As shown in Fig. \ref{fig_Tr}~(a), for $x_0<x_0^*\approx0.58$, $\langle
T_{x_0}(\mu,r)\rangle$ exhibits a global minimum at a non-trivial
value of $\mu^*(x_0)$ and $r^*(x_0)$ which are both decreasing
functions of $x_0$ (see Fig. \ref{fig_mu_r}~(a) and (b)
respectively). In the limit $x_0 \to 0$, these curves converge to our
exact results in Eqs. (\ref{res_smallx0_intro}). In contrast, 
for $x_0>x_0^*$, our simulations show (see Fig. \ref{fig_Tr}(c)) that the minimum of
$\langle T_{x_0}(\mu,r)\rangle$ is instead reached at $\mu^*(x_0>x_0^*) = 0$. Fig. \ref{fig_Tr} (b) shows the critical case $x_0=x_0^*$. Quite remarkably, the
values of the optimal parameters $\mu^*(x_0)$ and $r^*(x_0)$ exhibit a
sharp discontinuity as $x_0$ crosses the critical value $x_0^* \approx
0.58$, as shown in Fig. \ref{fig_mu_r}. %

{\it The case $x_0>x_0^*$.} Our numerical simulations clearly indicate that, for $x_0>x_0^*$, $\mu^*(x_0>x_0^*) = 0$ but $r^*(x_0>x_0^*)$ is
a non trivial constant independent of $x_0$. We can actually compute this constant analytically as follows. Since $\mu^* = 0$, we analyze Eq. (\ref{PS}) in the limit $\mu \to 0$. In this limit, one can show that $\hat f(k)$ is almost flat in the $k$ space, with $\hat f(k) \approx e^{-1}$, valid for $e^{-1/\mu}\ll |k|\ll e^{1/\mu}$. Substituting $\hat f(k) \approx e^{-1}$ in Eq. (\ref{PS}), we find that $\tilde{Q}_{x_0}(r=0,z)$ takes the simple
expression:
\begin{equation}\label{tildeQ_musmall}
\lim_{\mu \to 0}\tilde{Q}_{x_0}(r=0,z)=\frac{1}{\sqrt{1-z}\sqrt{1-{z}/{e}}} \;,
\end{equation}
which is independent of $x_0$. From Eq. (\ref{tildeQ_musmall}) together with Eq. (\ref{Eq:Tr_main}) one obtains 
\begin{equation}
\lim_{\mu \to 0}\langle T_{x_0}(\mu,r) \rangle=\frac{1}{\sqrt{r}\sqrt{1-{(1-r)}/{e}}-r} \;.
\label{Eq:mu0limit}
\end{equation}
As a function of $r$, $\langle T_{x_0}(\mu\to 0,r) \rangle$ in (\ref{Eq:mu0limit}) has a unique minimum at the optimal value $r^*_>$ given in Eq. (\ref{rstar_mu0}). Substituting $r=r^*_>$ in Eq. (\ref{Eq:mu0limit}) gives the optimal value of the MFPT 
\begin{equation}\label{tstar}
T^*(x_0>x_0^*) =  \frac{2\sqrt{e}(\sqrt{e}+\sqrt{e-1})}{(e-1))} =5.6794\ldots \;.
\end{equation}
%
\begin{figure}[ht]
\includegraphics[width=\linewidth]{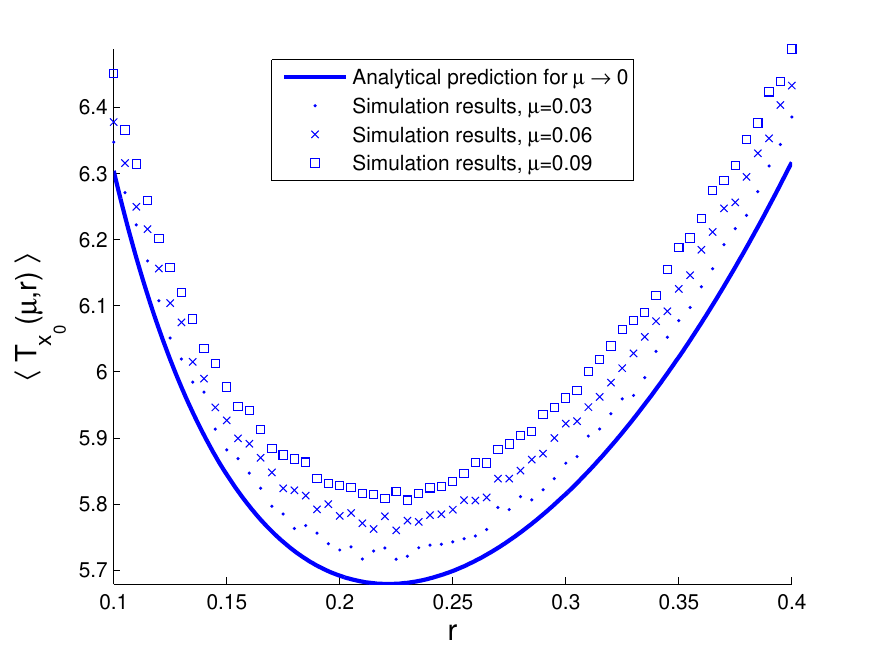}
\caption{\label{Fig:small_mu}$\langle T_{x_0}(\mu,r) \rangle$ vs. $r$ - comparison between numerical results for small $\mu$ and analytical prediction for $\mu \to 0$.}
\end{figure}
In Fig. \ref{Fig:small_mu}, we show a plot of $\langle T_{x_0}
(\mu,r)\rangle$ for $x_0 = 1 > x_0^*$, computed numerically, as a
function of $r$ and for different small values of $\mu$. This plot
confirms that, as $\mu \to 0$, the numerical data do converge to our
exact results in (\ref{Eq:mu0limit}) and (\ref{tstar}). 

Interestingly, our numerics also reveal the existence of a second special value $x_0^c\approx 0.56<x_0^*$, suggesting the following scenario as $x_0$
is increased from 0 to $\infty$. 
As $x_0$ increases, starting from $0$, $T_{x_0}(\mu,r)$ admits a single global minimum at $X_{\min}^{(1)} = (\mu^*(x_0) > 0,r^*(x_0))$ [see Fig. \ref{fig_Tr} (a)] until $x_0$ reaches the value $x_0 = x_0^c$. At this point a second local minimum appears at $X_{\min}^{(2)} = (\mu = 0, r_>^*)$. The value of $\langle T_{x_0}(\mu,r)\rangle$ at this local minimum at $X_{\min}^{(2)}$ however remains higher than the one at $X_{\min}^{(1)}$ until $x_0> x_0^*$. Therefore in this range when $x_0^c<x_0<x_0^*$ there are two competing local minima with $X_{\min}^{(1)}$ being the global minimum, and $X_{\min}^{(2)}$ being a metastable minimum [see Fig. \ref{fig_Tr} (b)]. When $x_0$ increases beyond $x_0^*$, then $X_{\min}^{(2)}$ becomes the global minimum [see Fig. \ref{fig_Tr} (c)]. This is then a typical scenario for a first order phase transition, as clearly illustrated in Fig. \ref{fig_mu_r}.  

This second value $x_0^c$ can actually be estimated analytically by studying the stability of the local minimum $X_{\min}^{(2)}$ starting from large $x_0$ where it is also a global minimum. We compute the sign of the
derivative of $\partial \langle T_r(x_0,\mu) \rangle/\partial \mu$
evaluated at $\mu=0$ and $r=r_>^*$ given in (\ref{rstar_mu0}).  A
straightforward computation, using Eqs. (\ref{PS}) and
(\ref{Eq:Tr_main}) shows that
\begin{equation}\label{sign}
\left. \mbox{sign} \left( \frac{\partial \langle T_r(x_0) \rangle}{\partial \mu} \right)\right|_{r=r^*_>,\mu=0} = {\rm sign}\left[\ln(x_0) + \gamma_E \right] \;,
\end{equation}
where $\gamma_E = 0.57721\ldots$ is the Euler constant. The slope does change sign from positive to negative as $x_0$ crosses from above the value $x_0^c = e^{-\gamma_E} = 0.56146\ldots < x_0^* \approx 0.58$. Our numerical estimate of $x_0^c$ is fully in agreement with the exact value $x_0^c = e^{-\gamma_E}$.

To summarize, for a searcher undergoing stable L\'evy jump processes with resetting in one-dimension, we showed that the MFPT to a fixed target at the origin has a rich phase diagram as a function of the L\'evy index $\mu$, the resetting probability $r$ and the starting position $x_0$. In particular, the optimal parameters $(\mu^*(x_0), r^*(x_0))$ that minimize the MFPT exhibit a surprising first-order phase transition at a critical value $x_0^*$. Our study leads to several open questions. For example, how generic is this first-order phase transition? Does it depend only on the tail or on other details of the jump distribution? Also, does this transition exist in higher dimensions and in presence of multiple searchers? These questions remain outstanding for future studies. 

\acknowledgements

SM and GS are supported by the ANR grant 2011-BS04-013-01 WALKMAT. SM, SS and GS are partially supported by the Indo-French Centre for the Promotion of Advanced Research under Project 4604-3. GS also acknowledges support from Labex PALM (Project RANDMAT). SM, SS and GS thank the Galileo Galilei Institute for TheoreticalPhysics, Florence, Italy for the hospitality and the INFN for partial support during the completion of this work.



\end{document}